\documentclass{article}
\usepackage{graphicx}
\usepackage[arrowdel]{physics}
\usepackage[rightcaption]{sidecap}
\graphicspath{ {./images/} }
\usepackage{cite}
\usepackage{float}
\usepackage{geometry}
\usepackage{authblk}
\usepackage{hyperref}

\providecommand{\keywords}[1]{\textbf{\textit{Keywords:}} #1}
\allowdisplaybreaks

\newcommand{\miktex}{\hbox{Mik\kern-.15em\TeX}}

\title{Polarization of charged spherical dielectric core-shell} 
\author[1]{A. Marin}
\author[1]{O. Ianc}
\author[1,a]{T. O. Cheche}
\affil[1]{University of Bucharest, Faculty of Physics, Măgurele, PO BOX MG11, 077125, Romania}
\affil[a]{Corresponding author email: \em{tiberius.cheche@unibuc.ro}}

\DeclareMathAlphabet\mathbfcal{OMS}{cmsy}{b}{n}
\begin{document}

\maketitle
\begin{abstract}
Two point charges are placed in a spherical dielectric  core-shell embedded in a dielectric environment, one of the charges being located in the core and the other in the shell. The core, shell, and environment are characterized by different dielectric constants. Analytical solutions for the electrostatic energy of the system and the polarization charge density at the core-shell and shell-environment interfaces are obtained by a classical approach.
\end{abstract}
%%%%%%%%%%%%%%%%%%%%%%%%%%%%%%%%
\section{Introduction}

Understanding the electro-optical properties of materials are of crucial importance in engineering components and devices which operate by the light-material interaction. The quantum  treatments of 
optical properties \cite{haug,cheche1,cheche2,cheche3} and spin Hall conductivity\cite{Murakami,cheche5,cheche6} in semiconductor structures, quantum approaches of the dynamics of photosynthetic system \cite{Nagarajan,cheche8}, the quantized versus the classical picture of crystal lattice oscillations \cite{Girvin,cheche9} all are based on the electrical properties of matter. Application of conservation laws is another powerful tool in modelling the classical \cite{voinea,cheche11,radu,cheche12,cheche13} or quantum motion. The electrostatic properties of dielectric materials are of interest for both academic learning and experimental research. Analytical solution for the polarization charge on a dielectric sphere due to the presence of a point charge is presented in the literature \cite{jackson,Raggi}.

In this work we consider a core-shell dielectric sphere embedded in a dielectric environment with two point charges located one in the core and the other in the shell.The core has the relative dielectric constant $\varepsilon_1$ and radius $r_0$, the shell has the relative dielectric constant $\varepsilon_2$ and core+shell radius is $R$, and the dielectric environment has the relative dielectric constant $\varepsilon_3$.  The goal of the work is two-fold, namely: i) calculus of the work classically necessary to assemble the charge distribution of two charges; ii) calculus of the polarization charge generated at the dielectric interfaces. The first task is motivated by the occurrence of the electron-hole pair after the light irradiation of type II core-shell semiconductor nanostructures, the so-called quantum dots (QDs), with the electron located in the shell and the hole located in the core in the fundamental quantum state. The second task once completed offers an intuitive picture on the charge induced by polarization and a result which can be compared with that obtained by a quantum density functional theory calculations \cite{Raggi}. The structure of the article is as follows. In Sec. 2 the potential generated by the two point charges is obtained for the dielectric core-shell. Section 3 provides the expression of the electrostatic energy of the system of the two point charges in dielectric in both Brus \cite{brus1983} and core-shell geometry. In Sec. 4, the surface charge induced by polarization by the two point charges in the and dielectric core-shell is obtained and graphically represented. Section 5 is dedicated to conclusions. 

\section{The potential}
This section presents a derivation for the potentials generated by each of the charges in the system. The point charge $q$ located in $\mathbf{s}_{i}$  at distance $s_{i}=|\mathbf{s}_{i}|$ to the center of core generates the charge density $\rho(\mathbf{r})=q \delta\left(\mathbf{r}-\mathbf{s}_{i}\right)$; $i=1,2$ for core, shell, respectively. Consequently, the Poisson equation for isotropic dielectric constant in each region
$
\Delta \phi=\rho /\varepsilon_{i}
$
becomes a Laplace equation excepting the points occupied by the charges. To describe the potential of the point charge we use expansion \cite{jackson}:
\begin{equation}
  \frac{1}{|\mathbf{r}-\mathbf{s}|}=\sum_{n=0}^{\infty} \frac{r_{<}^{n}}{r_{>}^{n+1}} P_{n}(\cos \theta)  
\end{equation}
and take into account the azimuthal symmetry of the problem in writing the solution of Laplace equation.
\begin{figure}[H]
\centering
\includegraphics[width=0.5\textwidth]{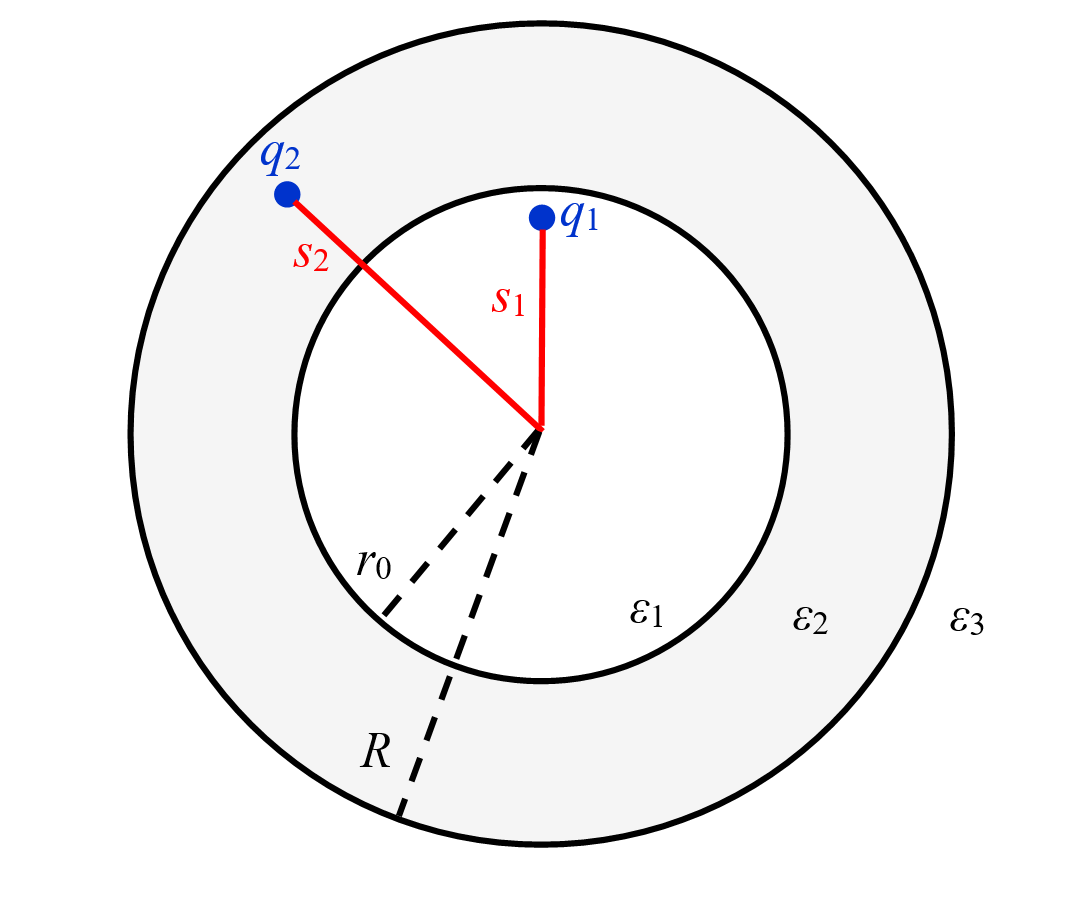}
\caption{Spherical core-shell dielectric. Cross-section through the center of the sphere. $\varepsilon_1$, $\varepsilon_2$, and $\varepsilon_3$ are relative dielectric constants.}
\label{configuration}
\end{figure}

%\item{} 
We can write the potential created by charge $q_{1}$ located at $\mathbf{s}_{1}$ on $z$ axis in $\mathbf{r}$ at distance $r=|\mathbf{r}|$ from origin and angle $\theta_{1}$ with respect to $z$ axis as follows:
\begin{subequations}\label{A1}
\begin{align}
\phi\left(s_{1}, r, \theta_{1}\right) &=\begin{cases}
&\sum_{n=0}^{\infty}\left[\frac{q_{1}}{4 \pi \varepsilon_{0} \varepsilon_{1} s_{1}}\left(\frac{r}{s_{1}}\right)^{n}+A_{1 n} r^{n}\right] P_{n}\left(\cos \theta_{1}\right), 0<r<s_{1} \\
&\sum_{n=0}^{\infty}\left[\frac{q_{1}}{4 \pi \varepsilon_{0} \varepsilon_{1} r}\left(\frac{s_{1}}{r}\right)^{n}+A_{1 n} r^{n}\right] P_{n}\left(\cos \theta_{1}\right), s_{1}<r<r_{0}
\end{cases}\label{A1_int}\\
\phi\left(s_{1}, r, \theta_{1}\right)=& \sum_{n=0}^{\infty}\left[B_{1 n} r^{n}+C_{1 n} r^{-n-1}\right]P_{n}\left(\cos \theta_{1}\right), r_{0}<r<R\label{A1_mid} \\
\phi\left(s_{1}, r, \theta_{1}\right) &=\sum_{n=0}^{\infty} D_{1 n} r^{-n-1} P_{n}\left(\cos \theta_{1}\right), r>R\label{A1_ext}
\end{align}
\end{subequations}
with $A_{1 n}, B_{1 n}, C_{1 n}, D_{1 n}$ constants. In eqs. (\ref{A1_int}) the first term of the sum is generated by the point charge and all the others in eqs. (\ref{A1}) are polarization charges induced by $q_{1}$. Boundary conditions, taking into account the orthogonality of the Legendre polynomials, are written as follows:
\begin{subequations}\label{A1_BC}
\begin{align}
\left.\frac{\partial \phi}{\partial \theta_{1}}\right|_{r_{0_{-}}}&=\left.\frac{\partial \phi}{\partial \theta_{1}}\right|_{r_{0_{+}}}, \\
\left.\varepsilon_{1} \frac{\partial \phi}{\partial r}\right|_{r_{0_{-}}}&=\left.\varepsilon_{2} \frac{\partial \phi}{\partial r}\right|_{r_{0_{+}}},\\
\left.\frac{\partial \phi}{\partial \theta_{1}}\right|_{R_{-}}&=\left.\frac{\partial \phi}{\partial \theta_{1}}\right|_{R_{+}}, \\
\left.\varepsilon_{2} \frac{\partial \phi}{\partial r}\right|_{R_{-}}&=\left.\varepsilon_{3} \frac{\partial \phi}{\partial r}\right|_{R_{+}}.    
\end{align}
\end{subequations}
From the system of equations (\ref{A1_BC}) in solving eqs. (\ref{A1}) one obtains
\begin{subequations}\label{A1_sol}
\begin{align}
&\phi\left(s_{1}, r, \theta_{1}\right)=\frac{q_{1}}{4 \pi \varepsilon_{0} \varepsilon_{1}} \frac{1}{\left|\mathbf{r}-\mathbf{s}_{1}\right|}+\sum_{n=0}^{\infty} \underbrace{\sigma_{1 n}^{A} \frac{s_{1}^{n}}{R^{2 n+1}}}_{A_{1 n}} r^{n} P_{n}\left(\cos \theta_{1}\right) , \quad 0<r<r_{0}\label{A1_sol_int}\\
&\phi\left(s_{1}, r, \theta_{1}\right)=\sum_{n=0}^{\infty}[\underbrace{\sigma_{1 n}^{B} \frac{s_{1}^{n}}{R^{2 n+1}}}_{B_{1 n}} r^{n}+\underbrace{\sigma_{1 n}^{C} s_{1}^{n}}_{C_{1 n}} \frac{1}{r^{n+1}}] P_{n}\left(\cos \theta_{1}\right) , r_{0}<r<R\label{A1_sol_mid}\\
&\phi\left(s_{1}, r, \theta_{1}\right)=\sum_{n=0}^{\infty} \underbrace{\sigma_{1 n}^{D} s_{1}^{n}}_{D_{1 n}} \frac{1}{r^{n+1}} P_{n}\left(\cos \theta_{1}\right), \quad r>R\label{A1_sol_ext}
\end{align}
\end{subequations}
with the constants $A_{1 n}, B_{1 n}, C_{1 n}, D_{1 n}$ given in Appendix.

Next, we write the potential created by charge $q_{2}$ located in $\mathbf{s}_{2}$ on (a new) $z$ axis in the field point $\mathbf{r}$ located at distance $r$ from the origin in the shell and angle $\theta_{2}$ with respect to $z$ axis as follows:
\begin{subequations}\label{A2}
\begin{align}
&\phi\left(s_{2}, r, \theta_{2}\right) =\sum_{n=0}^{\infty} A_{2 n} r^{n} P_{n}\left(\cos \theta_{2}\right), \quad 0<r<r_{0}\label{A2_int}\\
\begin{split}
&\phi\left(s_{2}, r, \theta_{2}\right)\\
&\hspace{0.7cm}=
\begin{cases}
&\sum_{n=0}^{\infty}\left[\frac{q_{2}}{4 \pi \varepsilon_{0} \varepsilon_{2} s_{2}}\left(\frac{r}{s_{2}}\right)^{n}+B_{2 n} r^{n}+C_{2 n} r^{-n-1}\right] P_{n}\left(\cos \theta_{2}\right), r_{0}<r<s_{2}\\
&\sum_{n=0}^{\infty}\left[\frac{q_{2}}{4 \pi \varepsilon_{0} \varepsilon_{2} r}\left(\frac{s_{2}}{r}\right)^{n}+B_{2 n} r^{n}+C_{2 n} r^{-n-1}\right] P_{n}\left(\cos \theta_{2}\right), s_{2}<r<R
\end{cases}\label{A2_mid}
\end{split}\\
&\phi\left(s_{2}, r, \theta_{2}\right)=\sum_{n=0}^{\infty} D_{2 n} r^{-n-1} P_{n}\left(\cos \theta_{2}\right), \quad r>R\label{A2_ext}
\end{align}
\end{subequations}
with $A_{2 n}, B_{2 n}, C_{2 n}, D_{2 n}$ constants. In eqs. (\ref{A2_mid}) the first term of the sum is generated by the point charge $q_{2}$ and all the others in eqs. (\ref{A2}) are polarization charges induced by $q_{2}$. Boundary conditions, taking into account the orthogonality of the Legendre polynomials give:
\begin{subequations}\label{A2_BC}
\begin{align}
&\left.\frac{\partial \phi}{\partial \theta_{2}}\right|_{r_{0_-}}=\left.\frac{\partial \phi}{\partial \theta_{2}}\right|_{r_{0_+}},
\\
&\left.\varepsilon_{1} \frac{\partial \phi}{\partial r}\right|_{r_{0_-}}=\left.\varepsilon_{2} \frac{\partial \phi}{\partial r}\right|_{r_{0_+}}, 
\\
&\left.\frac{\partial \phi}{\partial \theta_{2}}\right|_{R_-}=\left.\frac{\partial \phi}{\partial \theta_{2}}\right|_{R_+},
\\
&\left.\varepsilon_{2} \frac{\partial \phi}{\partial r}\right|_{R_-}=\left.\varepsilon_{3} \frac{\partial \phi}{\partial r}\right|_{R_+}.
\end{align}
\end{subequations}
From the system of equations (\ref{A2_BC}) in solving eqs. (\ref{A2}) one obtains:
\begin{subequations}\label{A2_sol}
\begin{align}
&\phi\left(s_{2}, r, \theta_{2}\right)=\sum_{n=0}^{\infty} \underbrace{\sigma_{2 n}^{A}\left(s_{2}\right) s_{2}^{-n-1}}_{A_{2 n}} r^{n} P_{n}\left(\cos \theta_{2}\right), 0<r<r_{0}\label{A2_sol_int}\\
\begin{split}
&\phi\left(s_{2}, r, \theta_{2}\right)
= \frac{q_{2}}{4 \pi \varepsilon_{0} \varepsilon_{2}} \frac{1}{\left|\mathbf{r}-\mathbf{s}_{2}\right|}\\
&\hspace{1cm}+\sum_{n=0}^{\infty}[\underbrace{\sigma_{2 n}^{B}\left(s_{2}\right) s_{2}^{-n-1}}_{B_{2 n}} r^{n}+\underbrace{\sigma_{2 n}^{C}\left(s_{2}\right) s_{2}^{n}}_{C_{2 n}} r^{-n-1}] P_{n}\left(\cos \theta_{2}\right), r_{0}<r<R
\end{split}\label{A2_sol_mid}\\
&\phi\left(s_{2}, r, \theta_{2}\right)=\sum_{n=0}^{\infty} \underbrace{\sigma_{2 n}^{D}\left(s_{2}\right) s_{2}^{-n-1}}_{D_{2 n}} r^{-n-1} P_{n}\left(\cos \theta_{2}\right), r>R\label{A2_sol_ext}
\end{align}
\end{subequations}
with the constants $A_{2n}, B_{2n}, C_{2n}, D_{2n}$ given in the Appendix.
In Appendix, the expressions obtained for the potentials of the two charges in eqs. (\ref{A1_sol}) and (\ref{A2_sol}) are checked by comparison with those from the literature for charge in a dielectric sphere embedded in a dielectric matrix.

\section{The electrostatic energy}
In this section we calculate the electrostatic energy $W$ of a system consisting of dielectrics of spherical geometry and two point charges incorporated in. By definition  $W$ is given by the work classically necessary to assemble the two charges in the system.
\subsection{Brus' problem}
\begin{figure}[h]
\centering
\includegraphics[width=0.4\textwidth]{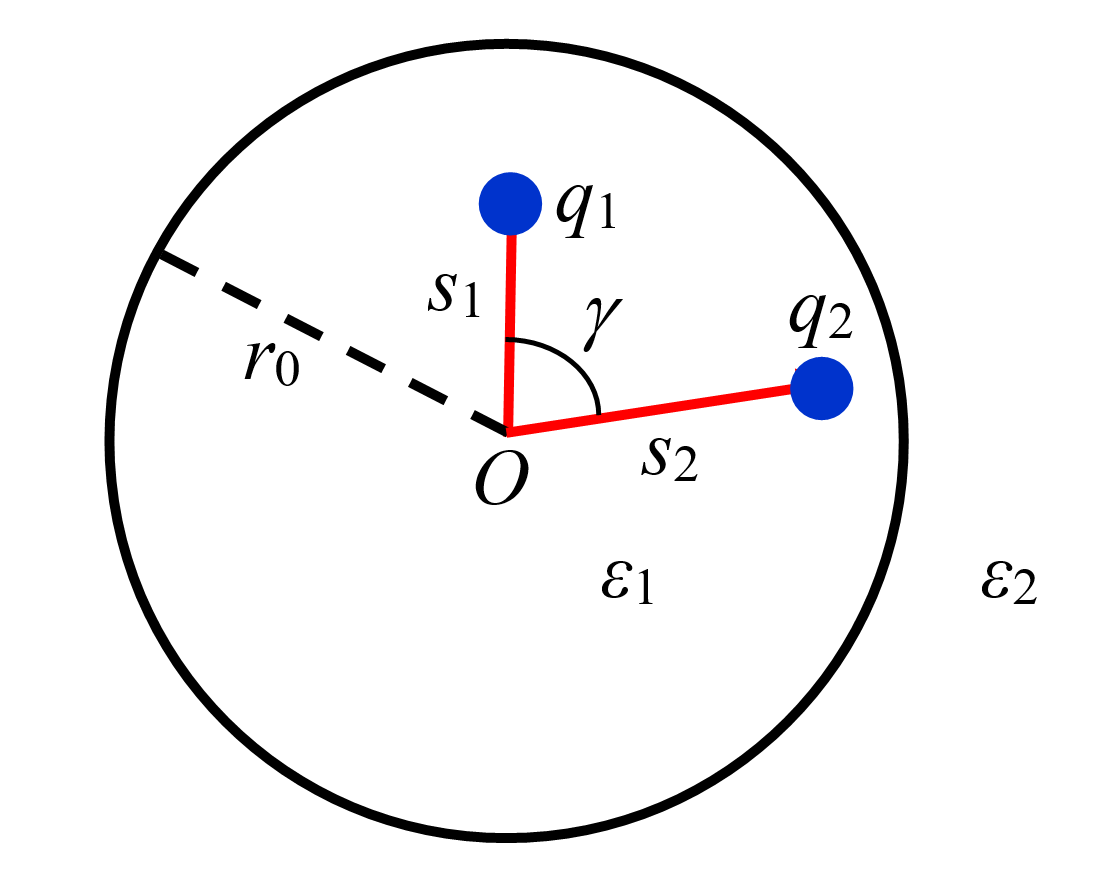}
\caption{Brus' problem: Point charges $q_1$, $q_2$ located inside the dielectric sphere of radius $r_0$. For computational reasons charges $q_1$, $q_2$ are imagined as being located on the surface of small spheres of radius $\delta$.
\label{brus_charges}
}
\end{figure}

Following the model of Brus \cite{brus1983}, we first calculate the work necessary to assemble two charges in a dielectric sphere of relative dielectric constant $\varepsilon_{1}$ and radius $r_0$ embedded in a dielectric of relative dielectric constant $\varepsilon_{2}$ as shown in Fig. \ref{brus_charges}.
First, we compute the work necessary to create one charge $q$ inside the sphere. For this we express the infinitesimal potential generated by an infinitesimal charge $dq$ put on the surace of a sphere of small  (atomic size) radius $\delta$ located inside the sphere:
\begin{figure}[h]
\centering
\includegraphics[width=0.4\textwidth]{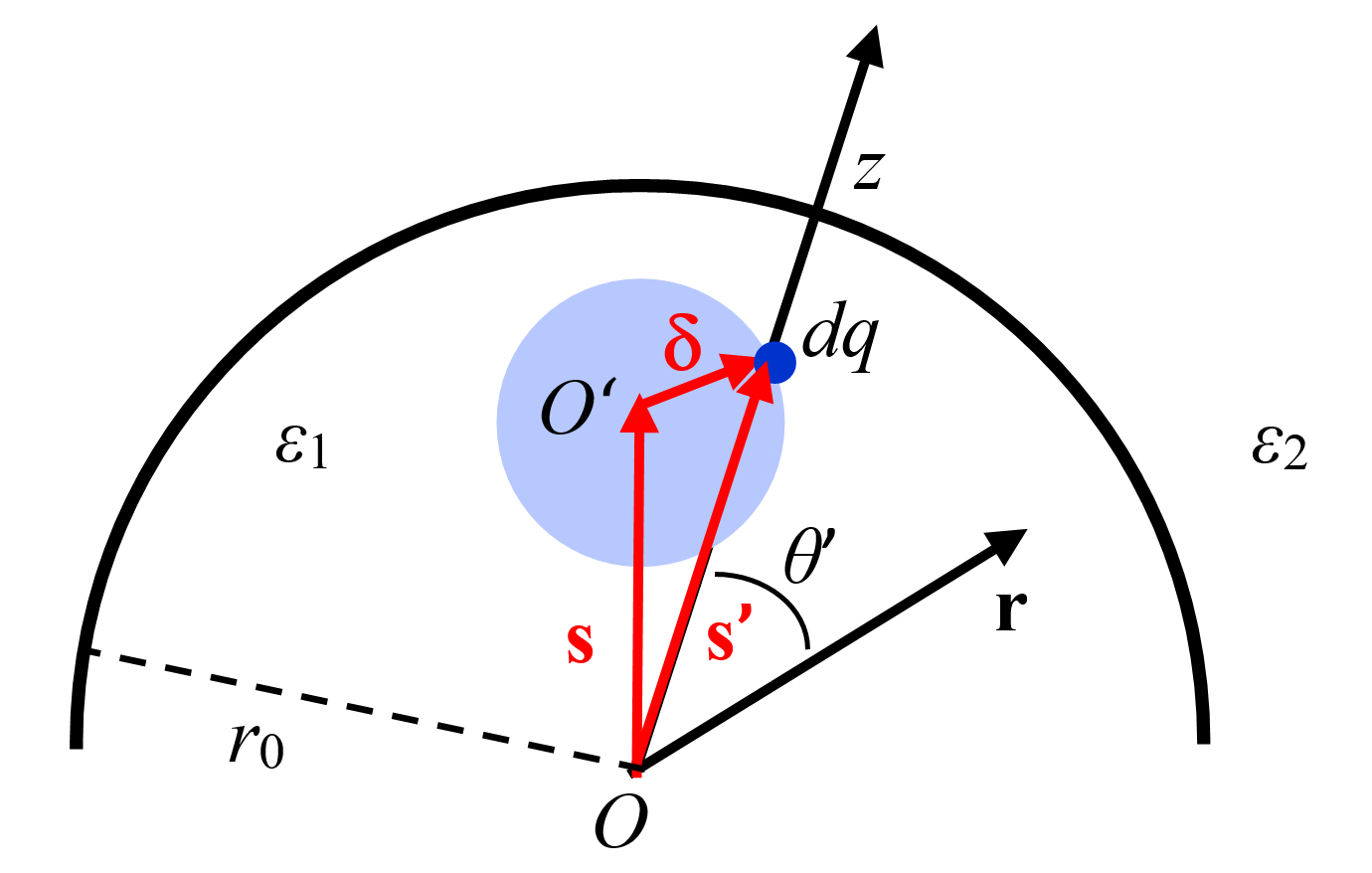}
\caption{Infinitesimal charge placed on the surface of sphere of small radius $\delta$.
\label{brus_image}
}
\end{figure}

\begin{equation}\label{1brus}
d \phi\left(s^{\prime}, r, \theta\right)=\frac{d q}{4 \pi \varepsilon_{0} \varepsilon_{1}} \frac{1}{\left|\mathbf{r}-\mathbf{s}^{\prime}\right|}+\sum_{n=0}^{\infty} \underbrace{d \sigma_{1 n}^{A} \frac{\left(s^{\prime}\right)^{n}}{r_{0}^{2 n+1}}}_{dA_{1n}} r^{n} P_{n}\left(\cos \theta^{\prime}\right), 0<r<r_0
\end{equation}
where $d \sigma_{1 n}^{A}=\frac{d q(\varepsilon-1)(n+1)}{\varepsilon_{1}(1+n+\varepsilon n)}$, and $\varepsilon=\varepsilon_{1} / \varepsilon_{2}$ (see Appendix, eq.\ref{App1} for the particular case $\varepsilon=1$). For $\delta$ of atomic size one has  $s \approx s^{\prime} \gg \delta$, the $z$ axis passes through $O^{\prime}$ (see Fig. \ref{brus_image}), and the integration over the small sphere surface in (eq. \ref{1brus}) results in:
\begin{equation}\label{2brus}
\phi(s, r, \theta)=\frac{q}{4 \pi \varepsilon_{0} \varepsilon_{1}} \frac{1}{|\mathbf{r}-\mathbf{s}|}+\sum_{n=0}^{\infty} \underbrace{\sigma_{1 n}^{A} \frac{s^{n}}{r_{0}^{2 n+1}}}_{A_{1 n}} r^{n} P_{n}(\cos \theta), 0<r<r_0
\end{equation}
where $\sigma_{1 n}^{A}=\frac{q(\varepsilon-1)(n+1)}{\varepsilon_{1}(1+n+\varepsilon n)}$ and $\theta$ is the angle between $\mathbf{r}$ and $\mathbf{s}$.
Then the work necessary to create the total charge $q$ is the work necessary to bring it from infinity to the small sphere surface centered at $\mathbf{s}$ (by a sequential infinitesimal process), that is:
\begin{equation}\label{work}
\begin{aligned}
&\tilde{W}_{q}=\int_{0}^{q} d q \phi\left(s, s+\delta, \delta \gamma\right) \stackrel{\operatorname{small} \delta}{\longrightarrow} \frac{q^{2}}{8 \pi \varepsilon_{0} \varepsilon_{1}} \frac{1}{\delta}+\sum_{n=0}^{\infty} \int_{0}^{q} d q \sigma_{ n}^{A} \frac{s^{2 n}}{r_{0}^{2 n+1}} \underbrace{P_{n}(\cos \delta \gamma)}_{\substack{\delta \gamma \rightarrow 0 \text { when } s \gg \delta}} \\
&=\underbrace{\frac{q^{2}}{8 \pi \varepsilon_{0} \varepsilon_{1}} \frac{1}{\delta}}_{W_{\text {Born }}}+\frac{q^{2}}{8 \pi \varepsilon_{0}} \underbrace{\sum_{n=0}^{\infty} \frac{(\varepsilon-1)(n+1)}{\varepsilon_{1}(1+n+\varepsilon n)} \frac{s^{2 n}}{r_{0}^{2 n+1}}}_{W_{q}}
\end{aligned}
\end{equation}
where $\delta$ is radius of the small sphere on which the infinitesimal charge $d q$ is added, and $\delta \gamma$ the angle between $\mathbf{s}\left(z\right.$ axis) and $\mathbf{s}+\boldsymbol{\delta}$.\

For the Brus' problem (see Fig. \ref{brus_charges}), the work necessary to bring the charge $q_{j}$ from infinity to $\mathbf{s}_j$ (which is located inside the dielectric sphere of radius $r_0$ of relative dielectric constant $\varepsilon_1$) in the field generated by the charge $q_{i}$ (also located inside the dielectric sphere of radius $r_0$) is calculated by using the potential from eq. (\ref{2brus}). Thus we have:
\begin{equation}\label{energy}
\begin{aligned}
& W_{i j}=\int_{0}^{q_{j}} d q_{j} \phi\left(s_{i}, s_{j}, \theta_{i j}\right)\\
=\quad & \frac{q_{i} q_{j}}{4 \pi \varepsilon_{0}}\left[\frac{1}{\varepsilon_{1}\left|s_{i}-s_{j}\right|}+\sum_{n=0}^{\infty} \frac{(\varepsilon-1)(n+1)}{\varepsilon_{1}(1+n+\varepsilon n)} \frac{s_{i}^{n} s_{j}^{n}}{r_{0}^{2 n+1}} P_{n}(\cos \theta_{i j})\right]
\end{aligned}
\end{equation}
with $\theta_{i j}=\gamma$ the angle between  $\mathbf{s}_i$ and  $\mathbf{s}_j$ and $i \neq j$.
Then, the work classically necessary to assemble the charge distribution (excluding the Born term) is
\begin{equation}\label{diel_en}
\begin{aligned}
W_{\text {diel }}=& W_{q_{1}}+W_{q_{2}}+\frac{1}{2}\left(W_{12}+W_{21}\right)=\underbrace{\frac{q_{1} q_{2}}{4 \pi \varepsilon_{0}} \frac{1}{\varepsilon_{1}\left|\mathbf{s}_{1}-s_{2}\right|}}_{W_{\text {charge }}}+\\
& \underbrace{\frac{1}{4 \pi \varepsilon_{0}} \sum_{n=0}^{\infty} \frac{(\varepsilon-1)(n+1)}{\varepsilon_{1}(1+n+\varepsilon n) r_{0}^{2 n+1}}\left[\frac{q_{1}^{2} s_{1}^{2 n}+q_{2}^{2} s_{2}^{2 n}}{2}+q_{1} q_{2} s_{1}^{n} s_{2}^{n} P_{n}(\cos \gamma)\right]}_{W_{\text {pol }}}
\end{aligned}
\end{equation}\
\subsection{Core-shell problem}
Next, we calculate the polarization energy for the charge distribution in spherical core-shell sketched in Fig. \ref{configuration}.

The energy $W_{q_{1}}$ necessary to assemble the charge $q_{1}$ at $\mathbf{s}_{1}$ and the energy $W_{q_{2}}$ necessary to assemble the charge $q_{2}$ at $\mathbf{s}_{2}$ are calculated by using the potential from eq. (\ref{A1_sol_int}) and respectively eq. (\ref{A2_sol_int}), with the recipe used to obtain eq. (\ref{work}). Thus, without the Born term, we have
\begin{equation}
\begin{aligned}
&W_{q_{1}}={\frac{q_{1}^{2}}{8 \pi \varepsilon_{0}} \sum_{n=0}^{\infty} \frac{1}{\widetilde{\varepsilon}_{1n}^{A}} \frac{s_{1}^{2 n}}{R^{2 n+1}}}
\end{aligned}
\end{equation}

\begin{equation}
\begin{aligned}
&W_{q_{2}}=\frac{q_{2}^{2}}{8 \pi \varepsilon_{0}} \frac{1}{s_{2}} \sum_{n=0}^{\infty}\left[\frac{1}{\tilde{\varepsilon}_{2 n}^{B}\left(s_{2}\right)}+\frac{1}{\tilde{\varepsilon}_{2 n}^{C}\left(s_{2}\right)}\right]
\end{aligned}
\end{equation}
The work necessary to bring the charge $q_{2}$ from infinity to $\mathbf{s}_2$ in the field generated by the charge $q_{1}$ located at $\mathbf{s}_1$ is calculated by using the potential from eq. (\ref{A1_sol_mid}) with the recipe used to obtain eq. (\ref{energy}). Thus we have:
\begin{equation}
\begin{aligned}
&W_{12}=\int_{0}^{q_{2}} d q \phi\left(s_{1}, s_{2}, \gamma\right)\\
&=\frac{q_{1} q_{2}}{4 \pi \varepsilon_{0}} \sum_{n=0}^{\infty}\left(\frac{1}{\tilde{\varepsilon}_{1 n}^{B}} \frac{s_{1}^{n} s_{2}^{n}}{R^{2 n+1}}+\frac{s_{1}^{n}}{\tilde{\varepsilon}_{l n}^{C}} \frac{1}{s_{2}^{n+1}}\right) P_{n}\left(\cos \gamma\right)
\end{aligned}
\end{equation}

The work necessary to bring the charge $q_{1}$ from infinity to $\mathbf{s}_1$ in the field generated by the charge $q_{2}$ located at $\mathbf{s}_2$ is calculated by using the potential from eq. (\ref{A2_sol_int}) with the recipe used to obtain eq. (\ref{energy}). Thus we have:
\begin{equation}
\begin{aligned}
&W_{21}=\int_{0}^{q 1} d q \phi\left(s_{2}, s_{1}, \gamma\right)=\frac{q_{2} q_{1}}{4 \pi \varepsilon_{0}} \sum_{n=0}^{\infty} \frac{s_{2}^{-n-1} s_{1}^{n}}{\tilde{\varepsilon}_{2 n}^{A}\left(s_{2}\right)} P_{n}\left(\cos \gamma\right)
\end{aligned}
\end{equation}
Similarly to the procedure used for eq. (\ref{diel_en}) one can obtain the electrostatic energy $W_{diel}$ of the core-shell system.
\section{The polarization charge}
\begin{figure}[h]
\centering
\includegraphics[width=0.6\textwidth]{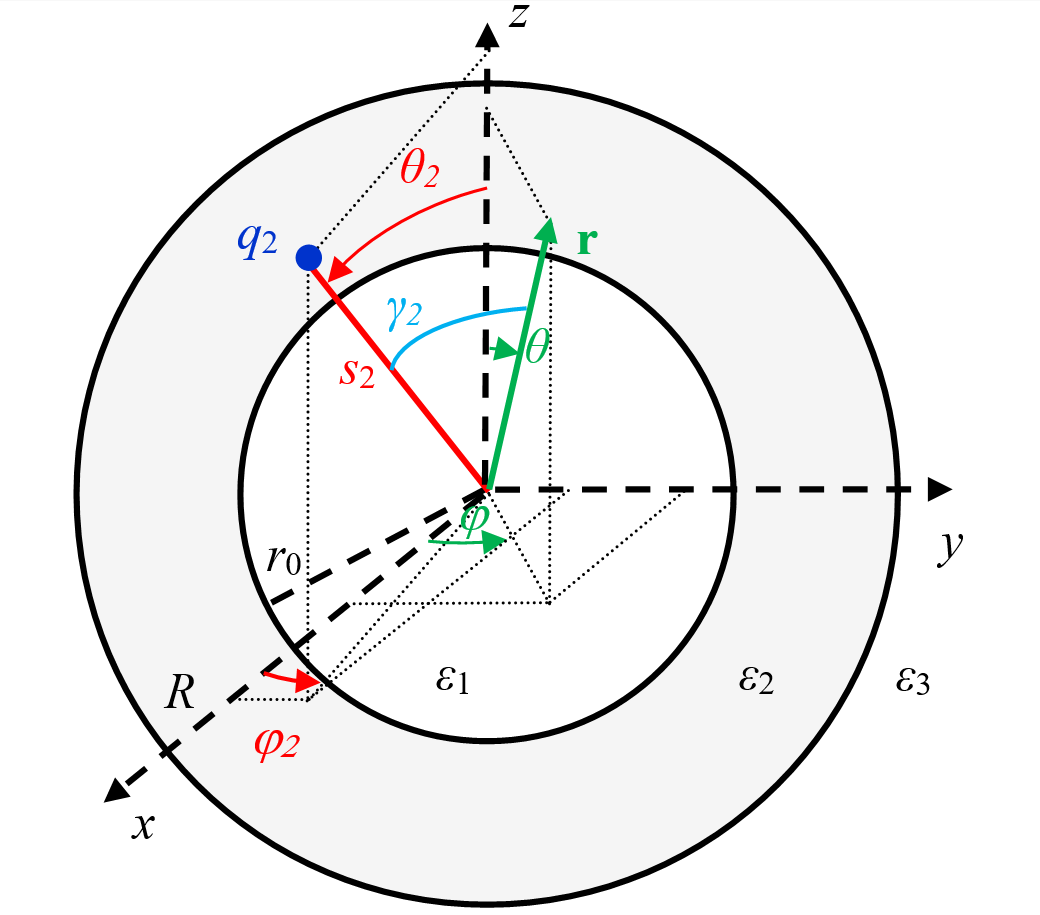}
\caption{The angular spherical coordinates used in the application of Legendre addition theorem. A similar choice of coordinates is employed for $q_1$ (not represented here for clarity reasons).}
\label{coordinates}
\end{figure}

In this section we compute the surface charge induced by polarization at the core-shell and shell-environment interfaces for the structure sketched in Fig. \ref{configuration}. 
By using the superposition principle, the electric potential in the structure can be obtained by adding the potentials generated by the two charges. Thus, by using the expressions obtained in eqs. (\ref{A1_sol}) and (\ref{A2_sol}) and the notations from Fig. \ref{coordinates}, the potential at the position vector $\mathbf{r}$ is

\begin{equation}
\begin{aligned}
&\phi\left(s_{1}, s_{2}, r, \theta_{1}, \theta_{2},\varphi_1,\varphi_2,\theta,\varphi\right) =\phi_1\left(s_{1}, r, \gamma_1\right)+\phi_2\left(s_{2}, r, \gamma_{2}\right)\\
&=\sum_{n=0}^{\infty} f_{1 n}\left(s_{1}, r\right) P_{n}\left(\cos \gamma_1\right)+\sum_{n=0}^{\infty}  f_{2 n}\left(s_{2}, r\right) P_{n}(\cos \gamma_2).    
\end{aligned}
\end{equation}
To calculate the charge induced by polarization we need the field that, accordingly to the piecewise potential expressions from eqs. (\ref{A1_sol}) and (\ref{A2_sol}), is defined as follows:
\begin{subequations} \label{E}
\begin{align}
\mathbf{E_{1L}(r)}&=-\boldsymbol{\nabla}\phi(\mathbf{r}),\quad s_{1}<r\le r_{0}\\
\mathbf{E_{12}(r)}&=-\boldsymbol{\nabla}\phi(\mathbf{r}),\quad r_{0}<r\le R\\
\mathbf{E_{13}(r)}&=-\boldsymbol{\nabla}\phi(\mathbf{r}),\quad r>R\\
\mathbf{E_{21}(r)}&=-\boldsymbol{\nabla}\phi(\mathbf{r}),\quad r\le r_{0}\\
\mathbf{E_{2S}(r)}&=-\boldsymbol{\nabla}\phi(\mathbf{r}),\quad r_0\le r \le s_{2}\\
\mathbf{E_{2L}(r)}&=-\boldsymbol{\nabla}\phi(\mathbf{r}),\quad s_{2}<r\le R\\
\mathbf{E_{23}(r)}&=-\boldsymbol{\nabla}\phi(\mathbf{r}),\quad r>R
\end{align}
\end{subequations}
The gradient, in spherical coordinates, can be obtained by writing the potential generated by $q_j$ as a function of variables $r,\theta,\varphi$ by using the Legendre addition theorem:
\begin{equation}
\begin{aligned}
 &P_{n}\left(\cos \gamma\right)=P_{n}\left(\cos \theta\right)P_{n}\left(\cos \theta_j\right)\\
 &+2\sum_{m=1}^{m=n} \frac{(n-m)!}{(n+m)!}P_{n}^{m}\left(\cos \theta\right) P_{n}^{m}\left(\cos \theta_j\right)\cos [m (\varphi-\varphi_j)],
 \label{legendre}
\end{aligned}    
\end{equation}
%with $j\in \{1,2\}$.
with $j=1, 2$.
The polarization (dipole moment volume density) is defined as $\mathbf{P}=\varepsilon_{0}\chi_{e}\mathbf{E}$, with $\chi_{e}=\varepsilon/\varepsilon_0-1$ the dielectric susceptibility. Since inside each dielectric $\boldsymbol{\nabla}{\cdot}\mathbf{E}=0$, except at the point charges, the polarization-surface charge density, $\sigma_{pol}=-\boldsymbol{\nabla}\cdot{\mathbf{P}}$, is also zero except both the point charges and the interfaces, where a jump in the $\chi_{e}$ value occurs. 
Then the polarization-surface charge density at the two interfaces
\begin{subequations}
\begin{align}
    \sigma_{pol}^{12}=-(\mathbf{P_{2}}-\mathbf{P_{1}})\cdot\mathbf{n},\\
    \sigma_{pol}^{23}=-(\mathbf{P_{3}}-\mathbf{P_{2}})\cdot\mathbf{n},
\end{align}
\end{subequations}
%with $\mathbf{n}$ the radial unit vector oriented towards the center of the sphere, is written as:
with $\mathbf{n}$ the radial unit vector oriented in the direction of increasing radius, is written as:
\begin{subequations}
\begin{align}
 \left[\chi_{e_2}(\mathbf{E_{12}}|_{r=r_{0{+}}}+\mathbf{E_{2S}}|_{r=r_{0{+}}})-\chi_{e_1}(\mathbf{E_{1L}}|_{r=r_{0{-}}}+\mathbf{E_{21}}|_{r=r_{0{-}}})\right]\cdot\mathbf{n}=-\frac{\sigma_{pol}^{12}}{\varepsilon_0},\\
  \left[\chi_{e_3}(\mathbf{E_{13}}|_{r=R_{+}}+\mathbf{E_{23}}|_{r=R_{+}})- \chi_{e_2}(\mathbf{E_{12}}|_{r=R_{-}}+\mathbf{E_{2L}}|_{r=R_{-}})\right]\cdot\mathbf{n}=-\frac{\sigma_{pol}^{23}}{\varepsilon_0}.
\end{align}
\end{subequations}
When computing the dot product, we are left only with the radial component of $\mathbf{E}$, so applying eqs. (\ref{E}) gives:
\begin{subequations}
\begin{align}
    -\chi_{e_2}\left(\frac{\partial \phi_{12}}{\partial r}|_{r=r_{0_{+}}}+\frac{\partial \phi_{2S}}{\partial r}|_{r=r_{0_{+}}}\right)+\chi_{e_1}\left(\frac{\partial \phi_{1L}}{\partial r}|_{r=r_{0_{-}}}+\frac{\partial \phi_{21}}{\partial r}|_{r=r_{0_{-}}}\right)=-\frac{\sigma_{pol}^{12}}{\varepsilon_0},\label{sigma12}\\
     -\chi_{e_3}\left(\frac{\partial \phi_{13}}{\partial r}|_{r={R_{+}}}+\frac{\partial \phi_{23}}{\partial r}|_{r={R_{+}}}\right)+\chi_{e_2}\left(\frac{\partial \phi_{12}}{\partial r}|_{r={R_{-}}}+\frac{\partial \phi_{2L}}{\partial r}|_{r={R_{-}}}\right)=-\frac{\sigma^{23}_{pol}}{\varepsilon_0},\label{sigma23}
\end{align}
\end{subequations}

\begin{figure}[h]
\includegraphics[width=.24\linewidth]{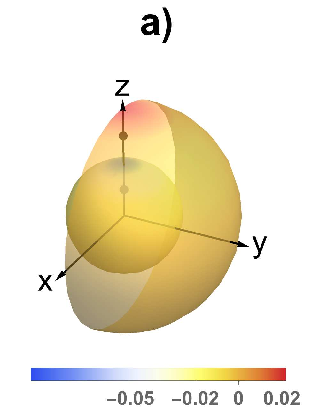}
\includegraphics[width=.24\linewidth]{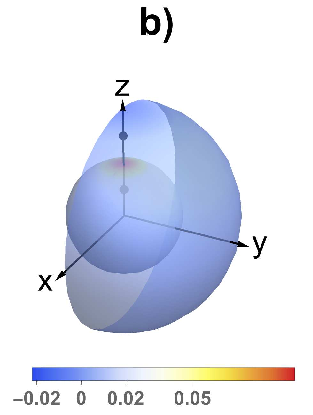}
\includegraphics[width=.24\linewidth]{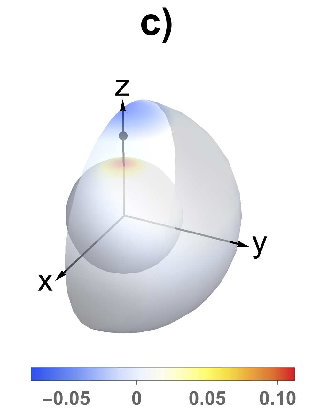}
\includegraphics[width=.24\linewidth]{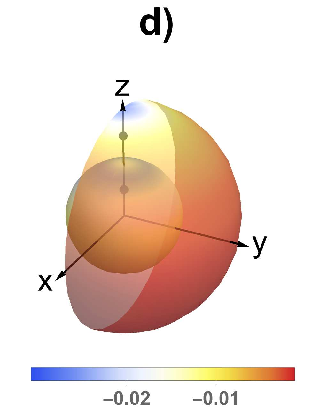}
\label{figs}
\caption{Polarization charge density for $r_0=1, R=2, s_1=0.5,\,s2=1.5,\,\varepsilon_1=2,\,\varepsilon_2=3,\,\varepsilon_3=4$, $\theta_1=\theta_2=0$ and: a) $q_1=1,\,q_2=-1$; b) $q_1=-1,\,q_2=1$; c) $q_1=0,\,q_2=1$; d) $q_1=1,\,q_2=1$ (SI units and rationalized to $\varepsilon_0$ dielectric constants).} 
\label{}
\end{figure}

One can check that the system behaves symmetrically when reflected on $Oxy$ plane. Indeed, by applying the known formula $P_n^m(-x)=(-1)^{n+m}P_n^m(x)$ to eq. (\ref{legendre}), it can be easily deduced that the reflection $\theta\to\pi-\theta,\,\theta_j\to\pi-\theta_j$ leaves $P_n(\cos\gamma)$ unchanged. In Fig. \ref{figs} we represented the polarization charge density for several values and positions of the point charges obtained from eqs. (\ref{sigma12}, \ref{sigma23}). Other cases are shown in Appendix. The code used for our numerical simulations is available  \href{https://github.com/andrei-marin29/polarization_charge}{here}.
\section{Conclusions}

For the system formed by two point charges placed in a spherical dielectric core-shell embedded in a dielectric environment, one of the charges being located in the core and the other in the shell, we obtained the the position-dependent potential, the electrostatic energy, and the polarization charge located at the core-shell and shell-environment interfaces. To obtain the analytical solutions we used a Legendre polynomial series expansion of the potential and appropriate boundary conditions. Some calculation checks are implemented. We checked the expression for the potential for particular cases with \cite{boettcher}, the expressions for the potential energy with \cite{brus1983} and the surface polarization charge density formulae by reflecting on $xy$ plane. Both classical expressions of electrostatic energy and polarization charge can be used for the comparison of classical results with results obtained by using microscopic approaches as those used in refs.,  e.g., \cite{Raggi, Lian, Ohshima}. The electrostatic energy we obtained can be quantized and used for an increased accuracy description of the electronic energy structure  and optical spectra simulation in core-shell QDs\cite{cheche1}.
\vspace{15mm}

\section*{Appendix}
Next, the details required to obtain the complete solutions are provided as follows.\\
\renewcommand\thefigure{A\arabic{figure}} 
\setcounter{figure}{0}
\setcounter{equation}{0}
\renewcommand{\theequation}{A.\arabic{equation}}
The constants $A_{1n}, B_{1n}, C_{1n}, D_{1n}$ that occur in eqs. (\ref{A1_sol}) are as follows:
\begin{subequations}\label{A1_App}
\begin{align}
&A_{1n}=\frac{q_{1}}{4
\pi \varepsilon_{0} \widetilde{\varepsilon}_{1 n}^{A}} \frac{s_{1}^{n}}{R^{2 n+1}}=\sigma_{1 n}^{A} \frac{s_{1}^{n}}{R^{2 n+1}},\\
\begin{split}
&\frac{1}{\widetilde{\varepsilon}_{1n}^{A}}=(1+n)\\
&\hspace{0.4cm}\times\frac{\left[\varepsilon_{1}+n\left(\varepsilon_{1}+\varepsilon_{2}\right)\right]\left(\varepsilon_{2}-\varepsilon_{3}\right)+\left(R / r_{0}\right)^{1+2 n}\left(\varepsilon_{1}-\varepsilon_{2}\right)\left[\varepsilon_{3}+n\left(\varepsilon_{2}+\varepsilon_{3}\right)\right]}{D n(n) \varepsilon_{1}}, 
\end{split}\\
%&B_{1}(n)\left[\sim \operatorname{meter}^{-n-1}\right]=\frac{q_{1}}{4
&B_{1n}=\frac{q_{1}}{4
\pi \varepsilon_{0} \widetilde{\varepsilon}_{1 n}^{B}} \frac{s_{1}^{n}}{R^{2 n+1}}=\sigma_{1 n}^{B} \frac{s_{1}^{n}}{R^{2 n+1}},\\
&\frac{1}{\widetilde{\varepsilon}_{1 n}^{B}}=\frac{(1+2
n)(1+n)\left(\varepsilon_{2}-\varepsilon_{3}\right)}{D n(n)}, \\
&C_{1n}=\frac{q_{1}
s_{1}^{n}}{4 \pi \varepsilon_{0} \widetilde{\varepsilon}_{1 n}^{C}}=\sigma_{1 n}^{C} s_{1}^{n},\\
&\frac{1}{\widetilde{\varepsilon}_{1 n}^{C}}=\frac{(1+2
n)\left[\varepsilon_{3}+n\left(\varepsilon_{2}+\varepsilon_{3}\right)\right]}{D n(n)}, \\
&D_{1n}=\frac{q_{1}
s_{1}^{n}}{4 \pi \varepsilon_{0} \widetilde{\varepsilon}_{1 n}^{D}}=\sigma_{1 n}^{D} s_{1}^{n},\\
&\frac{1}{\widetilde{\varepsilon}_{1 n}^{D}}=\frac{(1+2 n)^{2} \varepsilon_{2}}{D n(n)},\\
\begin{split}
&Dn(n)=n(1+n)\left(\varepsilon_{1}-\varepsilon_{2}\right)\left(\varepsilon_{2}-\varepsilon_{3}\right)\left(r_{0} / R\right)^{1+2 n}\\
&\hspace{1.2cm}+\left[\varepsilon_{2}+n\left(\varepsilon_{1}+\varepsilon_{2}\right)\right]\left[\varepsilon_{3}+n\left(\varepsilon_{2}+\varepsilon_{3}\right)\right].
\end{split}
\end{align}
\end{subequations}\

The constants $A_{2n}, B_{2n}, C_{2n}, D_{2n}$ that occur in eqs. \ref{A2_sol} are as follows:
\begin{subequations}\label{A2_App}
\begin{align}
&A_{2n}
=\frac{q_{2}}{4 \pi \varepsilon_{0} \widetilde{\varepsilon}_{2 n}^{A}\left(s_{2}\right)} s_{2}^{-n-1}=\sigma_{2 n}^{A}\left(s_{2}\right) s_{2}^{-n-1}, \\
&\frac{1}{\widetilde{\varepsilon}_{2 n}^{A}}=\frac{(1+2 n)\left[\varepsilon_{3}+n\left(\varepsilon_{2}+\varepsilon_{3}\right)+\left(\varepsilon_{2}-\varepsilon_{3}\right)(1+n)\left(s_{2}^{1+2 n} / R\right)^{1+2 n}\right]}{D n(n)}, \\
&B_{2n}
=\frac{q_{2}}{4 \pi \varepsilon_{0} \widetilde{\varepsilon}_{2 n}^{B}\left(s_{2}\right)} s_{2}^{-n-1}=\sigma_{2 n}^{B}\left(s_{2}\right) s_{2}^{-n-1}, \\
\begin{split}
&\frac{1}{\widetilde{\varepsilon}_{2 n}^{B}}=\left(\varepsilon_{2}-\varepsilon_{3}\right)(1+n)\\
&\hspace{0.7cm}\times\frac{{\left(\varepsilon_{2}-\varepsilon_{1}\right) n\left(r_{0} / R\right)^{1+2n}+\left[\varepsilon_{2}+\left(\varepsilon_{1}+\varepsilon_{2}\right) n\right]\left(s_{2} / R\right)^{1+2 n}}}{\varepsilon_{2} D n(n)},
\end{split}
\\
&C_{2n}
=\frac{q_{2}}{4 \pi \varepsilon_{0} \widetilde{\varepsilon}_{2 n}^{C}\left(s_{2}\right)} s_{2}^{-n-1} r_{0}^{1+2 n}=\sigma_{2 n}^{C}\left(s_{2}\right) s_{2}^{-n-1} r_{0}^{1+2 n},  \\
&\frac{1}{\widetilde{\varepsilon}_{2 n}^{C}}=\frac{\left(\varepsilon_{1}-\varepsilon_{2}\right)\left\{-n\left[\varepsilon_{3}+\left(\varepsilon_{2}+\varepsilon_{3}\right) n\right]+n(1+n)\left(\varepsilon_{3}-\varepsilon_{2}\right)\left(s_{2} / R\right)^{1+2 n}\right\}}{\varepsilon_{2} D n(n)}, \\
&D_{2n}
=\frac{q_{2}}{4 \pi \varepsilon_{0} \widetilde{\varepsilon}_{2 n}^{D}\left(s_{2}\right)} s_{2}^{-n-1} r_{0}^{1+2 n}=\sigma_{2 n}^{D}\left(s_{2}\right) s_{2}^{-n-1} r_{0}^{1+2 n},  \\
&\frac{1}{\widetilde{\varepsilon}_{2 n}^{D}}=\frac{(1+2 n)\left\{-n\left(\varepsilon_{1}-\varepsilon_{2}\right)+\left[\varepsilon_{2}+\left(\varepsilon_{1}+\varepsilon_{2}\right) n\right]\left(s_{2} / r_{0}\right)^{1+2 n}\right\}}{D n(n)}.
\end{align}
\end{subequations}
\

The expressions obtained for the potentials of the two charges in eqs. (\ref{A1_sol}) and (\ref{A2_sol}) are checked by comparison with those from ref.\cite{boettcher} for one point charge in a spherical dielectric embedded in a dielectric matrix as follows.\\
If in eq. (\ref{A1_App}) one sets $\varepsilon_2=\varepsilon_3=1$, we obtain:
\begin{subequations}\label{App1}
\begin{align}
    &A_{1n}=\frac{(n+1) q_1 \left(\varepsilon
    _1-1\right) r_0^{-2 n-1} s_1^n}{4 \pi \varepsilon _0 \varepsilon
    _1 \left(n \varepsilon _1+n+1\right)},\\
    &B_{1n}=0,\\
    &C_{1n}=\frac{(2 n+1) q_1 s_1^n}{4 \pi
    \varepsilon _0 \left(n \varepsilon _1+n+1\right)},\\
    &D_{1n}=\frac{(2 n+1) q_1 s_1^n}{4 \pi
    \varepsilon _0 \left(n \varepsilon _1+n+1\right)}.
\end{align}
\end{subequations}
which are similar to the expressions from ref.\cite{boettcher}, p. 83.
\\
If we set $\varepsilon_1=\varepsilon_2=\varepsilon_3$, the space becomes a homogeneous dielectric and eqs. (\ref{A1_App}) transform to:
\begin{subequations}
\begin{align}
    &A_{1n}=0,\\
    &B_{1n}=0,\\
    &C_{1n}=\frac{q_1 s_1^n}{4 \pi \varepsilon _0
    \varepsilon _1},\\
    &D_{1n}=\frac{q_1 s_1^n}{4 \pi \varepsilon _0
    \varepsilon _1}.
\end{align}
\end{subequations}
which introduced in eqs. (\ref{A1_sol}) express the potential of a point charge placed at distance $s_1$ to the origin of system of reference.\\
If in eqs. (\ref{A2_App}) one sets $\varepsilon_2=\varepsilon_1=1$,  we obtain:
\begin{subequations}
\begin{align}
    &A_{2n}=\frac{1}{4\pi\varepsilon_0}\frac{(2 n+1)q_2
   s_2^{-n-1}}{\varepsilon_1n+n+1},\\
   &B_{2n}=0,\\
   &C_{2n}=-\frac{1}{4\pi\varepsilon_0}\frac{(\varepsilon_1-1) n q_2 r_0^{2 n+1}
   s_2^{-n-1}}{\varepsilon_1 n+n+1},\\
   &D_{2n}=\frac{1}{4\pi\varepsilon_0}q_2 s_2^n-\frac{1}{4\pi\varepsilon_0}\frac{ (\varepsilon_1-1) n
   q_2r_0^{2 n+1} s_2^{-n-1}}{\varepsilon_1
   n+n+1}.
\end{align}
\end{subequations}
which are similar to the expressions from ref. \cite{boettcher}, p. 86.\\
The setting $\varepsilon_3=\varepsilon_2=\varepsilon_1$ in eqs. (\ref{A2_App}) implies:
\begin{subequations}
\begin{align}
    &A_{2n}=\frac{1}{4\pi\varepsilon_0}\frac{ q_2
    s_2^{-n-1}}{\varepsilon_1},\\
    &B_{2n}=0,\\
    &C_{2n}=0,\\
    &D_{2n}=\frac{1}{4\pi\varepsilon_0}\frac{ q_2 s_2^n}{\varepsilon_1}.
\end{align}
\end{subequations}
which introduced in eqs. (\ref{A2_sol}) express the potential of a point charge placed at distance $s_2$ to the origin of system of reference.\

\begin{figure}[H]
\includegraphics[width=.24\linewidth]{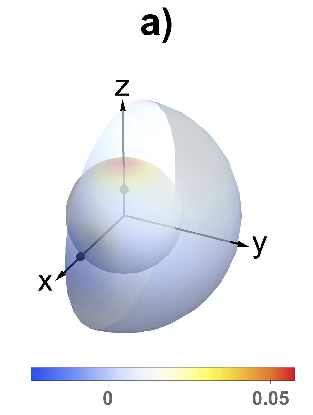}
\includegraphics[width=.24\linewidth]{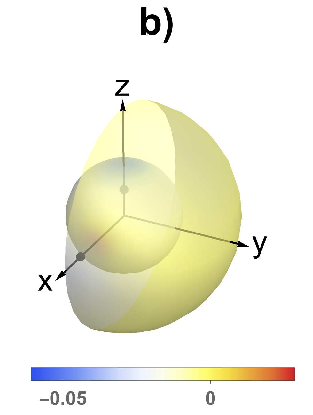}
\includegraphics[width=.24\linewidth]{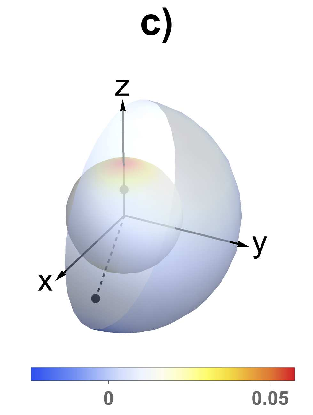}
\includegraphics[width=.24\linewidth]{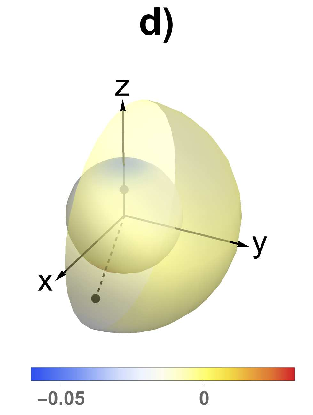}
%\caption{Polarization charge distribution for (rationalized units with $\varepsilon_0=1$) $r_0=1,\, R=2\, s_1=0.5,\,s2=1.5,\,\varepsilon_1=2,\,\varepsilon_2=3,\,\varepsilon_3=4$, $\theta_1=\varphi_1=\varphi_2=0$ and: a) $\theta_2=\pi/2,\,q_1=-1,\,q_2=1$; b) $\theta_2=\pi/2,\,q_1=1,\,q_2=1$; c) $\theta_2=3\pi/4,\,q_1=-1,\,q_2=1$; d) $\theta_2=3\pi/4,\,q_1=1,\,q_2=1$.}
%\end{figure}

%Fig. \ref{figs_app} shows additional configurations for which polarization charge distribution is computed.

\caption{Polarization charge density for $r_0=1, R=2, s_1=0.5,s2=1.5,\varepsilon_1=2,\varepsilon_2=3,\varepsilon_3=4$ and: a) $\theta_2=\pi/2,\,q_1=-1,\,q_2=1$; b) $\theta_2=\pi/2,\,q_1=1,\,q_2=1$; c) $\theta_2=3\pi/4,\,q_1=-1,\,q_2=1$; d) $\theta_2=3\pi/4,\,q_1=1,\,q_2=1$ (SI units and rationalized to $\varepsilon_0$ dielectric constants).}
%\label{A1}
\label{figs_app}
\end{figure}

In Fig. \ref{figs_app} we represented the polarization charge density for additional values and positions of the point charges obtained from eqs. (\ref{sigma12}, \ref{sigma23}).

%%%%%%%%%%%%%%%%%%%%%%%%%%%%%%%%

\end{document}